\documentclass[12pt]{article}
\usepackage{epsfig}
\def\be{\begin{equation}}
\def\ee{\end{equation}}
\newcommand{\bea}{\begin{eqnarray}}
\newcommand{\eea}{\end{eqnarray}}

\newcommand{\bp}{{\bf p}}
\newcommand{\bbe}{{\bf e}}

\newcommand{\Qb}{\overline{Q}}

\def\Pom{{\bf I\!P}}

\def\lsim{\mathrel{\rlap{\lower4pt\hbox{\hskip1pt$\sim$}}
    \raise1pt\hbox{$<$}}}         

\def\gsim{\mathrel{\rlap{\lower4pt\hbox{\hskip1pt$\sim$}}
    \raise1pt\hbox{$>$}}}         

\title{Production of spin-3 mesons in diffractive DIS}

\date{}
\author{F. Caporale$^{1}$, I.P. Ivanov$^{1,2}$\thanks{E-Mail:igivanov@cs.infn.it}\\
  {\normalsize $^1$ INFN, Gruppo Collegato di Cosenza, Italy}\\
  {\normalsize $^2$ Sobolev Institute of Mathematics,  Novosibirsk, Russia}}

\begin{document}
\maketitle

\begin{abstract}
We calculate the amplitudes of $J^{PC}=3^{--}$ meson production in diffractive DIS within 
the $k_t$-factorization approach,
with a particular attention paid to the $\rho_3(1690)$ meson.
We find that at all $Q^2$ the $\rho_3(1690)$ production cross section 
is 2--5 times smaller than the $\rho(1700)$ production cross section,
which is assumed to be a pure $D$-wave state. Studying $\sigma_L$ and $\sigma_T$
separately, we observe domination of $\rho_3$ in $\sigma_L$ and 
domination of $\rho(1700)$ in $\sigma_T$ and offer an explanation 
of this behavior in simple terms.
We also find very strong contributions --- sometimes even domination ---
of the $s$-channel helicity violating amplitudes. 
The typical color dipole sizes probed in $\rho_3$ production
are shown to be larger than those in the ground state $\rho$ production,
and the energy dependence of $\rho_3$ cross section turns out to be much flatter
than the $\rho$ production cross section. 
All the conclusions about the relative behavior of $\rho_3(1690)$ and $\rho(1700)$ mesons 
are numerically stable against variations of input parameters.
\end{abstract}

\section{Introduction}

Diffractive production of vector mesons (VM) in DIS $\gamma^* p \to Vp$ ($V = \rho,\phi,J/\psi$ etc.)
is a very active field of research (see recent review \cite{review} and references therein). 
Studying the $Q^2$-behavior of the VM production cross sections, 
one can learn about the transition from soft to hard regimes in strong interactions, while their
energy dependence reveals the Regge properties of the Pomeron exchange.
The rich set of possible helicity amplitudes $\gamma(\lambda_\gamma) \to V(\lambda_V)$
allows one to study the spin properties of the reaction and 
to test the $s$-channel helicity conservation (SCHC).

The main focus of this research has been on the production of the grounds state mesons,
while diffractive production of excited states did not enjoy much attention.
Perhaps, the most studied case so far was the production of radially excited mesons $V(2S)$.
Remarkable consequences of the presence of a node in the radial wave function
described in \cite{2s-th}  were nicely confirmed by H1 measurements
of diffractively produced $\psi(2S)$ \cite{psi2s-ex}.

Similar experimental studies of excited $\rho$ mesons are expected to be 
even more rewarding.
First, diffractive production of excited $\rho$ mesons
probes the dipole cross section at larger dipole sizes than the
production of ground states. For example, in analysis of \cite{dosch}
dipole sizes up to 2 fm were important. Such a unique sensitivity of these reactions
to soft diffraction can help understand the phenomenon of 
saturation, which is now a hot topic of debates (see \cite{saturation} and references therein). 
Another handle offered by diffractive production of $\rho'$
is their possible help in resolving the long standing puzzle of the radial/orbital excitation
assignment, as well as a possibility of a hybrid component in the $\rho(1450)$ 
and $\rho(1700)$ mesons, see \cite{4mesons}.

Diffractive production of excited $\rho'$ mesons
has been observed for a long time in a number of fixed target experiments 
with relatively high energies.
Diffractive production of $\rho'(1600)$ was reported in $\pi^+\pi^-$ 
\cite{aston1980} and $4\pi$ \cite{atkinson1985} final states
(for reanalysis of these data in terms of $\rho(1450)$ 
and $\rho(1700)$ mesons and for 
references to earlier experiments at lower energies, see \cite{tworhoprimes}). 
These states were also studied in a recent Fermilab experiment E687 \cite{E687}
both in $2\pi$ and $4\pi$ channels. 
Finally, the OMEGA Collaboration has succeeded to measure 
diffractive photoproduction cross section 
of the $\rho_3(1690)$ (known then as $g(1690)$ meson) 
via the $a_2(1320)\pi \to \eta\pi^+\pi^-$ diffractive final state \cite{omega1986}.
However, all these experiments gave only the value of the photoproduction cross section,
and no energy dependence, $Q^2$ dependence, or helicity structure
of the reaction was studied. This gap was partially closed by the 
H1 measurements of $\rho'$ electroproduction at $4 < Q^2 < 50$ GeV$^2$ \cite{H1},
but due to low statistics the results presented had large errorbars.

Diffractive production of excited states has not received too much attention
also from theory. Early theoretical discussions were limited to 
vector dominance models and its off-forward upgraded versions, see \cite{tworhoprimes}.
The pQCD based calculations of diffractive production of the ground state vector mesons
were developed in mid-90's and were almost immediately extended 
to the case of radial excitations, since the principal effect there is the presence 
of the node in the radial wave function \cite{2s-th}.
However, for a long time no microscopic calculation of the orbitally 
excitated vector mesons was available.

The situation was aggravated by an understanding that production of $D$-wave vector mesons
(as well as $J^{PC}=3^{--}$ mesons) should be suppressed by Fermi motion, 
as its radial wave function vanishes at the origin. 
This suppression was believed to be sufficiently strong and 
even prompted the authors of \cite{dosch} to consider diffractive $\rho(1450)$ and $\rho(1700)$
production neglecting in both cases the $D$-wave contributions altogether.
Only in \cite{in99} were the $S$-wave and $D$-wave vector meson
production amplitudes calculated within the $k_t$-factorization approach,
however at that time the absence of convenient parameterizations 
of the unintegrated gluon density --- the key input quantity --- 
impeded numerical predictions.

During the last years, several fits to the unintegrated gluon density 
appeared \cite{dgsf}, \cite{in2000}.
This allowed for the first estimates  \cite{phd} of the purely 
$D$-wave VM production cross sections, which showed that at small to moderate $Q^2$ 
the production rates of the $D$-wave and $2S$ $\rho'$ states are roughly comparable.
This was not surprising, since similar conclusion was drawn in \cite{tworhoprimes} 
during the famous splitting the $\rho'(1600)$ into $\rho(1450)$
and $\rho(1700)$. For a more detailed analysis of the $D$-wave vector meson
production in $k_t$-factorization, see \cite{Dwave}.

The approach developed in \cite{in99} can be applied also to the diffractive 
production of spin-3 mesons. The $D$-wave vector meson and the spin-3 meson
can be viewed as spin-orbital splitting partners and can be described within the same
formalism.  The only modification required to proceed from $D$-wave VM to spin-3 meson is that 
of the $q\bar q$ coupling to the final meson:
\be
\bar u\, {\cal D}^\mu\, u \cdot V^*_\mu  \to \bar u\, {\cal C}^{\mu\nu\rho}\, u 
\cdot T^*_{\mu\nu\rho}\,,\label{coupling3}
\ee
where $V_\mu$ and $T_{\mu\nu\rho}$ are the polarization vector for spin-1 and polarization tensor 
for spin-3 mesons, respectively. The spinorial structure ${\cal D}^\mu$ determined in \cite{in99} 
corresponds to the pure $D$-wave vector meson, 
while the structure ${\cal C}^{\mu\nu\rho}$ to be found
should correspond to $D$-wave spin-3 meson. 

In this paper we report the first microscopic 
derivation of the $J^{PC}=3^{--}$ production amplitudes 
within the $k_t$-factorization approach.
We focus on the $\rho_3(1690)$ production and give predictions 
on the $Q^2$ and $W$ dependence of the cross sections,
on the $\sigma_L/\sigma_T$ decomposition, and on the role of $s$-channel helicity violating
amplitudes. We also compare the $\rho_3(1690)$ cross sections with those of $\rho(1700)$, 
which is assumed to be purely $D$-wave vector meson, 
and observe a number of remarkable distinctions.
These should prove useful in disentangling these two mesons in experiment,
especially in the case of low statistics.

The paper is organized as follows. In Section 2 we 
show how spin-3 mesons are described and
present analytic expressions of the spin-3 meson production amplitudes. 
In Section 3 we perform twist analysis of the amplitudes in the forward case,
compare the results with those of $D$-wave vector mesons,
and discuss the effects of $s$-channel helicity violating amplitudes,
which will appear in the non-forward scattering. 
In Section 4 we present numerical results for $\rho_3$ and compare them with
the corresponding ground state $\rho$ and $\rho(1700)$ cross sections.
In Section 5 we discuss the typical color dipole sizes probed in $\rho_3$
production, compare our results with experimental data available
and comment on future possibilities. Finally, in Section 6 
we draw our conclusions.

\section{Amplitudes of spin-3 meson production}

\subsection{Kinematics and notation}

We use the usual notation for kinematical variables. $Q^2$ is 
the photon's virtuality, $W$ is the total center-of-mass energy of the $\gamma^*p$ collision.
The momentum transfer from proton to photon is denoted by $\Delta_\mu$ and at high energies 
is almost purely transverse: $-\Delta^2 = |t| \approx |t'| = \vec \Delta^2$. 
The transverse vectors (orthogonal to the $\gamma^*p$ collision axis)  
will be always labelled by an arrow.

Diffractive production of meson $V$ with mass $m_V$ can be treated in the lowest
Fock state approximation 
as production of the corresponding $q\bar q$ pair of invariant mass $M\not = m_V$, 
which is then projected, at the amplitude level, onto the final state.
Within the leading log${1 \over x}$ accuracy the higher Fock states are reabsorbed into
the evolution of the unintegrated gluon density (or color dipole cross section).
A typical diagram to be calculated (see \cite{review}) contains the valence quark loop,
with integration over the quark transverse momentum $\vec k$ 
and its fraction of photon's lightcone momentum $z$,
and the uppermost gluon loop, with the integration over transverse momentum $\vec \kappa$. 
A convenient choice is to assign momentum $\vec k + z\vec\Delta$ to the quark and 
$-\vec k +(1-z)\vec\Delta$ to the antiquark, which ensures that 
even at non-zero $\vec \Delta$ the $q\bar q$ invariant mass is 
$M^2 = {m^2 + \vec k^2 \over z(1-z)}$. 
It is also convenient to consider the relative $q\bar q$ momentum 
$p^\mu \equiv (k_q - k_{\bar q})^\mu/2$ in the $q\bar q$ pair rest frame, 
where it reduces to the 3-momentum $\bp = (\vec k, k_z)$ 
with $k_z \equiv {2z-1 \over 2}M$. 
Such 3-dimensional vectors, which always refer to the $q\bar q$ rest frame, 
will be given in bold.

Finally, throughout the text the ground state vector mesons 
(which will be always understood as $1S$ states)
will be generically labelled by $V$
or $V_{1S}$, the pure $D$-wave vector mesons will 
be labelled by $V_D$, while the $J^{PC}=3^{--}$ states
of the same quarkonium will be marked as $V_3$.
When we speak of $D$-wave mesons,
we will always assume $D$-wave {\em vector mesons},
not spin-3 mesons (although in $V_3$ the $q\bar q$ pair
also sits in the $D$-wave). In order to avoid excessive subscripts,
$m_V$ will refer to the mass of the meson being discussed. 
In particular, for the $\rho$ system
we will speak of $\rho_3$ and $\rho_D$. 
The former refers to the physical $\rho_3(1690)$ state,
while the identification of the latter with $\rho(1700)$ is clearly model dependent
and is done only for purposes of comparison.
The effects of $S$-wave admixture in $\rho(1700)$ are considered in \cite{Dwave}.

\subsection{Description of a spin-3 meson}

A spin-3 particle is described with the rank-3 polarization tensor $T^{\mu\nu\rho}$,
which must be symmetric and traceless in any pair of Lorentz indices. States
with any given helicity can be written as simple combinations
of well-known polarization vectors $e^\mu_\lambda$, $\lambda = +1,\,0,\,-1$ 
(see below more on $e^\mu_0$):  
\bea
&&T_{+3}^{\mu\nu\rho} = e^\mu_+ e^\nu_+ e^\rho_+\,,\qquad
T_{+2}^{\mu\nu\rho} = {1\over \sqrt{3}}\{e^\mu_+ e^\nu_+ e^\rho_0\}\,,\nonumber\\
&&T_{+1}^{\mu\nu\rho} = {1\over \sqrt{15}}\left(2\{e^\mu_+ e^\nu_0 e^\rho_0\}
+ \{e^\mu_+ e^\nu_+ e^\rho_-\}\right)\,,\quad
T_{0}^{\mu\nu\rho} = {1\over \sqrt{10}}\left(2e^\mu_0 e^\nu_0 e^\rho_0
+ \{e^\mu_+ e^\nu_0 e^\rho_-\}\right)\,,
\label{polarization}
\eea
where curly brackets denote symmetrization.

To construct the spinorial structure ${\cal C}^{\mu\nu\rho}$, recall first that
both quark and antiquark spinors can be treated on-mass-shell (see details in \cite{phd}).
Therefore,  ${\cal C}^{\mu\nu\rho}$ can be contructed from two independent structures: 
$\gamma^\mu p^\nu p^\rho$ and $p^\mu p^\nu p^\rho$ (symmetrization over indices is assumed). 
When constructing ${\cal C}^{\mu\nu\rho}$, we should make sure that it represents purely $L=2$
state without admixture of $L=4$ wave, 
which is in complete analogy with construction of $S$ and $D$-wave vector mesons
performed in \cite{in99}. Note that we explicitly rely on the lowest Fock state 
approximation of the meson; if higher Fock states are taken into account, such simple picture
is lost.

${\cal C}^{\mu\nu\rho}$ in (\ref{coupling3}) can be constructed in a most trasnparent way
in the non-relativistic case.
Instead of the $D$-wave vector meson structure $\varphi^\dagger\, D^{ij}\sigma^j\, \varphi$,
where $D^{ij} \equiv 3p^ip^j - \delta^{ij}\bp^2$, we have now
$\varphi^\dagger\, D^{ij}\sigma^k\, \varphi$.
Recall that this structure will be contracted with polarization tensor $T^{ijk}$
so that one should not worry about symmetrization. Since this polarization tensor
is traceless, the coupling simplifies to $\varphi^\dagger\,\sigma^k\, \varphi\, p^i p^j$,
which is just a tensor product of the $S$-wave coupling for vector meson 
and the $p^i p^j$ term. We do not have to keep track of the overall normalization factors,
since they can be always absorbed in the definition of the radial wave function.

Returning to the fully relativistic case, we can now write the spinorial structure
for spin-3 meson coupling
\be
\bar u\, {\cal C}^{\mu\nu\rho}\, u \cdot T_{\mu\nu\rho} = 
\bar u\, {\cal S}^{\mu}p^\nu p^\rho\, u \cdot T_{\mu\nu\rho} \equiv 
\bar u\, {\cal S}^{\mu}\, u \cdot \tau_{\mu}\,,\quad \tau_\mu \equiv  
T_{\mu\nu\rho}p^\nu p^\rho\,.
\label{C}
\ee
As the last form of this coupling shows,
the spin-3 amplitudes are easy to construct once
we know the amplitudes for the $S$-wave vector meson.

Given the spinorial structure, one can now calculate various amplitudes
with spin-3 meson. In such calculations, the radial wave function $\psi(\bp^2)$
will appear, whose normalization is
\be
1 = {N_c \over(2\pi)^3}\int d^3 \bp\, 4M |\psi_3(\bp^2)|^2 \cdot \left(-\tau^\mu \tau_\mu^*\right)
= {N_c \over(2\pi)^3} {2 \over 15} \int d^3 \bp\, 4M \bp^4 |\psi_3(\bp^2)|^2\,.
\label{norm3}
\ee
Apart from coefficient $1/15$, this normalization condition
coincides with the $D$-wave vector meson normalization condition. 

We underline that our approach to describing polarization states 
of the final meson is explicitly rotationally-invariant.
We use {\em identical} radial wave functions for all the polarization 
states of the final meson, and in this way 
the normalization condition (\ref{norm3}) 
holds for an arbitrary polarization state of the spin-3 meson.
An important part of the rotation-invariant description is
that the transversity condition must be imposed at the level of
the $q\bar{q}$ pair. This leads to the concept of the {\it
running longitudinal} polarization vector $e^\mu_{0}(M)$, 
such that it is orthogonal to the 4-momentum 
of the on-mass shell $q\bar{q}$ pair with invariant mass $M$
rather than the momentum of the final meson.
The calculations with the fixed longitudinal
polarization vector often found in literature
break the rotation invariance.
In technical terms, the fixed polarization vector leads to 
a mixing of the longitudinal spin-1 state and spin-0 states.

Potential models suggest that the radial wave functions
of the spin-orbital partners should be very similar.
One can assume, as a starting approximation, 
that their shapes are identical. This assumption leads to
\be
\psi_3(\bp^2) = \sqrt{15}\,\psi_D(\bp^2)\,,\label{wf}
\ee
which will be useful for comparison of spin-3 and $D$-wave vector meson production.

\subsection{Generic amplitudes}

A generic form of the helicity amplitudes 
$\gamma^*(\lambda_\gamma) \to V_3(\lambda_3)$ is
\be
Im\, A_{\lambda_3;\lambda_\gamma} =  W^2 {c_V \sqrt{4\pi\alpha_{em}} \over 4 \pi^2} 
\int {dzd^2\vec k\over z(1-z)}  \int {d^2\vec \kappa \over \vec \kappa^4}
\alpha_s\,{\cal F}(x_1,x_2,\vec \kappa,\vec\Delta)\cdot
I^{(3)}_{\lambda_3;\lambda_\gamma}\cdot\psi_3(\bp^2)\,.\label{ampl}
\ee
Here $c_V$ is the flavor-dependent average charge
of the quark, the argument of the strong coupling constant 
$\alpha_s$ is max$[z(1-z)(Q^2+M^2),\vec\kappa^2]$, 
and ${\cal F}(x_{1},x_{2},\vec \kappa,\vec\Delta)$ 
is the skewed unintegrated gluon distribution,
with $x_1 \not = x_2$ being the fractions of the proton's momentum
carried by the uppermost gluons.
The appearance of skewed (or generalized) parton distributions
is characteristic for scattering processes that change the mass/virtuality
of the projectile \cite{GPD}, see also recent reviews \cite{GPDreview}. In the case of meson
production, its use is important due to $x_2 \ll x_1$
and has been incorporated in the collinear factorization
approach \cite{GPD,GPDcollinear} as well as in the factorization approach with 
non-zero transverse momenta of quarks taken into account \cite{goloskokov}.
In the $k_t$-factorization approach, the skewness is transferred to
the unintegrated distributions.

The integrands for spin-3 mesons $I^{(3)}_{\lambda_3;\lambda_\gamma}$ 
can be written in terms of the corresponding integrands for vector mesons:
\bea
I^{(3)}_{+3;\lambda_\gamma} &=& I_{+;\lambda_\gamma} (k_+^*)^2\,,\label{i3}\\ 
I^{(3)}_{+2;\lambda_\gamma} &=& {1 \over \sqrt{3}}\left(2I_{+;\lambda_\gamma} k_z k_+^* + 
 I_{0;\lambda_\gamma} (k_+^*)^2 \right)\,,\label{i2}\\ 
I^{(3)}_{+1;\lambda_\gamma} &=& {1 \over \sqrt{15}}\left[
(2k_z^2 - \vec k^2)I_{+;\lambda_\gamma} + 4 k_z k_+^* I_{0;\lambda_\gamma}
+ (k_+^*)^2 I_{-;\lambda_\gamma}\right]\,,\label{i1}\\ 
I^{(3)}_{0;\lambda_\gamma} &=& {1 \over \sqrt{10}}\left[
(2k_z^2 - \vec k^2)I_{0;\lambda_\gamma} + 2 k_z k_-^* I_{+;\lambda_\gamma}
+ 2 k_z k_+^* I_{-;\lambda_\gamma}\right]\,,\label{i0}
\eea
where we used shorthand notation:
\be
k_\pm \equiv - (p_\mu e_\pm^\mu) = \bp\cdot\bbe_\pm = - k^*_\mp\,,\quad 
k_z \equiv  - (p_\mu e_L^\mu) =  \bp\cdot\bbe_0\,.\nonumber
\ee
Note that both $p_\mu$ and $e_\lambda^\mu$ depend on $\vec\Delta$, 
still this dependence of their scalar product vanishes due to the Lorentz invariance. 
The integrands $I_{\lambda_i;\lambda_\gamma}$ are:
\bea
I_{0;0} &=& 4QMz^2 (1-z)^2\left[1+ {(2z-1)^2 \over 4z(1-z)}{2m \over M+2m}\right]\, 
\Phi_2\,,\label{i00}\\[2mm]
I_{+;+} &=& m^2 \Phi_2 + [z^2 + (1-z)^2] \Phi_{1+}k_+^* + 
{m\over M+2m}\left[\vec k^2 \Phi_2 - (2z-1)^2  \Phi_{1+}k_+^*\right]\,,\label{ipp}\\[2mm]
I_{-;+} &=& 4z(1-z) \left[1+ {(2z-1)^2 \over 4z(1-z)}{2m \over M+2m}\right] k_+ \Phi_{1+} 
- {2m \over M+2m}k_+^2 \Phi_2 \,,\label{ipm}\\[2mm]
I_{0;+} &=& -4z(1-z)\left[1+ {(2z-1)^2 \over 4z(1-z)}{2m \over M+2m}\right] k_z \Phi_{1+}
+ {2m \over M+2m} k_z k_+ \Phi_2\,,\label{ip0}\\[2mm]
I_{+;0} &=& -4z(1-z) {Q \over M} k_z k_+^* {M \over M+2m} \Phi_2\,.\label{i0p}
\eea
The integrands for helicity $-1$ can be obtained from those with $+1$ 
by replacement of $k_+ \to k_-$ and $\Phi_{1+} \to \Phi_{1-}$ 
(with no extra minus sign that would appear only at the level of amplitudes!).
Here function $\Phi_2$ describes transition of virtual photon into the 
$q \bar{q}$ states with $\lambda_q+\lambda_{\bar q}=\lambda_{\gamma^*}$,
whereas $\vec\Phi_1$ describes transition of transverse photons into the $q\bar{q}$ states with 
$\lambda_q+\lambda_{\bar q}=0$, in which the helicity of the photon is carried 
by the orbital angular momentum in the $q\bar{q}$ state:
\bea
\Phi_2 &=& -{1\over (\vec r+\vec\kappa)^2 + \Qb^2} -{1 \over
(\vec r-\vec\kappa)^2 + \Qb^2} + 
{1 \over (\vec r + \vec\Delta/2)^2 + \Qb^2} + {1 \over (\vec r -
\vec\Delta/2)^2 + \Qb^2} \, ,\nonumber\\
\vec\Phi_1 &=&
-{\vec r + \vec\kappa \over (\vec r+\vec\kappa)^2 +
\Qb^2} -{\vec r - \vec\kappa \over (\vec r-\vec\kappa)^2
+ \Qb^2} + {\vec r + \vec\Delta/2 \over (\vec r +
\vec\Delta/2)^2 + \Qb^2} + {\vec r - \vec\Delta/2 \over
(\vec r - \vec\Delta/2)^2 + \Qb^2} \,,
\nonumber
\eea
where $\vec r \equiv \vec k + (2z-1)\vec \Delta/2$ and $\Qb^2 \equiv z(1-z)Q^2 + m^2$.

\section{Large $Q^2$, $m_V^2$ analysis}

The above expressions can be integrated numerically. However, before describing these results
it is useful to study analytically the case where both $Q^2$ and $m_V^2$ are much larger than
any soft scale, while $Q^2/m_V^2$ can be arbitrary. In this approximation one expands
the hard scale $\Qb^2$ around $\Qb_0^2 \equiv {1 \over 4}(Q^2 + m_V^2)$,
as well as performs expansion in powers of small Fermi motion of the $q\bar q$ pair, 
$k_z^2, \vec k^2 \ll m_V^2$. We will call this approximation the ``twist'' expansion.  
We start with the forward case, $\vec \Delta = 0$, where only the $s$-channel 
helicity conserving amplitudes with $\lambda_3 = \lambda_\gamma$ survive,
and find the ratio $\sigma_L/\sigma_T$ as well as relation between
$V_3$ and $V_D$ production cross sections. After this, we
discuss the role of helicity violating amplitudes.

\subsection{Twist expansion for the forward case}

We consider the two non-zero integrands in this approximation (\ref{i1}) and (\ref{i0}) and 
note that after $d\Omega_{\bp}$ angular averaging all terms in each of these expressions 
give comparable contributions, $\propto \bp^4\vec \kappa^2/\Qb_0^4$, 
differing only by numerical coefficients: 
\bea
I^{(3)}_{+;+} &=& {\vec \kappa^2 \over \Qb_0^4}\bp^4 \left[{3 \over 15}\left(1 + 
{8 \over 3}{M^2 \over Q^2 + M^2}\right) - {6\over 15} + {3\over 15}\right] 
={\vec \kappa^2 \over \Qb_0^4}\cdot {8\over 15}{\bp^4}{M^2 \over Q^2 +M^2}\,,\nonumber\\[2mm]
I^{(3)}_{0;0} &=& {Q\over M}\cdot{\vec \kappa^2 \over \Qb_0^4}\bp^4 
\left[{2 \over 15}\left(1 + {4M^2 \over Q^2 + M^2}\right) + {2\over 15}\right]
= {Q\over M}\cdot{\vec \kappa^2 \over \Qb_0^4}\cdot {4\over 15}\bp^4 
\left(1 + {2M^2 \over Q^2 + M^2}\right)\,.\nonumber
\eea
It is curious to note that the leading-twist contribution in
the transverse amplitudes vanishes, and one is left with the subleading term.
This cancellation does not occur in the longitudinal amplitude,
which leads to un abnormally large value of the ratio $\sigma_L/\sigma_T$:
\be
R_{LT} \equiv {\sigma_L \over \sigma_T}\cdot {m_V^2 \over Q^2} = 
{27 \over 8}\left(1 + {Q^2 \over 3 m_V^2}\right)^2 \gg 1\,.\label{rlt}
\ee
This must be confronted with $R_{LT} =1$ for the ground state mesons
and, even more remarkably, with $R_{LT} \ll 1$ for $D$-wave vector mesons,
evaluated within the same approximation. 

We stress that such a peculiar $Q^2$-dependence of the ratio $\sigma_L/\sigma_T$
(\ref{rlt}) is entirely due to the heavy-meson approximation we used.
Allowing for the longitudinal quark motion will restore the leading twist
contribution to the transverse amplitude. What is expected to remain, however, 
is the overall smallness of the transverse amplitude and, therefore,
a large numerical value of $R_{LT}$.

\subsection{Spin-3 vs. $D$-wave vector meson}

Within the twist expansion, the $V_D$ and $V_3$ production amplitudes
are proportional to $\int d^3 \bp\, \bp^4 \psi_D(\bp^2)$ 
and $\int d^3 \bp\, \bp^4 \psi_3(\bp^2)$, respectively. 
Assuming (\ref{wf}), one can relate the $V_D$ and $V_3$ production cross sections.
The results for the longitudinal and transverse cross sections, separately, are
\be
{\sigma_L^{(3)}\over \sigma_L^{(1)}} = 24\left({1 + {2m_V^2 \over Q^2 + m_V^2}\over 
1 - {8m_V^2 \over Q^2 + m_V^2}}\right)^2 \gg 1\,,\quad 
{\sigma_T^{(3)} \over \sigma_T^{(1)}} = 4\left({1\over 
1 + {15 \over 4}{Q^2+m_V^2 \over m_V^2}}\right)^2 \ll 1\,.\label{s31twist}
\ee
where we assumed the masses of the two states to be equal.
One sees that the longitudinal cross section is dominated by $V_3$ meson,
while the transverse one is dominated by $V_D$. In some sense,
these two mesons ``mirror'' each other: where $V_3$ is suppressed, $V_D$ dominates
and vice versa. 

This ``mirror'' behavior can be in fact understood in simple terms.
Consider, for instance, the $T\to T$ transition.
Note first that although the integrands for $V_D$ derived in \cite{in99} look very
differently from those of $V_3$, they can be written in a form similar to (\ref{coupling3}): 
$\bar u D^\mu u \cdot V_\mu \equiv \bar u S^\mu u \cdot D^{\mu\rho}\cdot V_\rho$,
where $D^{\mu\rho}\cdot V^\rho_+ = -{1\over 2}
\left[(2k_z^2 - \vec k^2)e_+^\mu - 6 k_zk_+ e_0^\mu + 6(k_+)^2 e_-^\mu\right]$.
This should be compared with $\tau^\mu_{+1} = {1\over \sqrt{15}}\left[(2k_z^2 - \vec k^2)e_+^\mu 
+ 4 k_zk_+ e_0^\mu + (k_+)^2 e_-^\mu\right]$.
The corresponding integrands $I^{(3)}_{+;+}$ and $I^{(1)}_{+;+}$ are
\be
\begin{array}{l}
I^{(3)}_{+;+} =  {1 \over \sqrt{15}} \left[(2k_z^2 - \vec k^2) I_{+;+}
+ 4 k_z k_+^* I_{0;+} + (k_+^{*})^2 I_{-;+}\right]\,;\\[3mm]
I^{(1)}_{+;+} = -{1\over 2} \left[(2k_z^2 - \vec k^2) I_{+;+}
- 6 k_z k_+^* I_{0;+} + 6(k_+^{*})^2 I_{-;+}\right]\,,
\end{array}
\label{reason}
\ee
The key point is the opposite signs in front of the second term.
It turns out that the contributions of all three terms in $I^{(1)}_{+;+}$
are of the same sign and of the same order of magnitude,
so that they interfere constructively in $V_D$ production.
In the case of $V_3$, they interfere destructively,  
which leads to suppressed $\sigma_T^{(3)}$.

For the longitudinal amplitude, the similar change of signs
strongly suppresses the result for $D$-wave mesons, 
enhancing it in the spin-3 case. We see
that there are good reasons to expect such a ``mirror'' behavior 
of $V_3$ and $V_D$ just on the basis of their spin-angular composition. 

\subsection{The role of $s$-channel helicity violation}

The approximate conservation of the $s$-channel helicity in diffractive reactions is due
to two reasons. First, in diffraction the helicity properties of the target and projectile
are uncorrelated, and in the forward case strict SCHC holds separately for the projectile 
and the target.
At $\vec\Delta\not = 0$ the $s$-channel helicity non-conserving (SCHNC) amplitudes depend 
on the momentum transfer as $|\vec\Delta|^{|\lambda_\gamma-\lambda_V|}$,
which makes them small within diffractive cone. Second, at high energy 
the helicity is conserved at the parton level, so in order to produce
helicity flip the transverse motion of constituent must come into play. 
This produces extra factors like $\vec k^2/M^2$, further suppressing helicity violation,
especially for heavy quarks.
Nevertheless, small violation of SCHC has been observed at HERA 
in the case of light vector mesons; at the amplitude level, 
its relative magnitude was estimated to be $\sim 10\%$ \cite{SCHNCHERA}.

In the case of $V_3$ production, the effect of SCHNC must be more important,
just as it was for $D$-wave vector meson production \cite{Dwave}.
The integration of the quadrupole term kills the leading contribution
to the SCHC amplitudes (\ref{i00}) and (\ref{ipp}),
and the SCHC amplitudes get the same suppression due to Fermi motion
as SCHNC ones.
Moreover, partial cancellation among several terms discussed above suppresses 
the $T \to T$ amplitude, while the helicity violating amplitudes
do not suffer such cancellation.
Finally, pay attention to numerical factors like $1/\sqrt{15}$ in the 
amplitude $A_{+1;+1}$, which are absent, 
for example, in the $A_{+3;+1}$ amplitude and take into account the large
number of various helicity violating amplitudes for spin-3 meson production.

Thus, one can anticipate that the helicity violating amplitudes can generate
a significant portion of the overall cross section. One should not even be surprised
to see them dominate in the transverse cross section, especially at small $Q^2$.
Therefore, the above twist analysis is meant only to guide the eye and should not be
used for quantitative discussion.

\section{Numerical study}

In this section we present numerical results for the particular case of $\rho_3$ production.

\subsection{Input}

In order to integrate (\ref{ampl}) numerically, one needs to specify 
models for the unintegrated gluon density and the meson wave function. 
We related the skewed unintegrated gluon density with non-zero
momentum transfer to the forward unintegrated gluon density by
\be
{\cal F}\left(x_1,x_2,\vec\kappa+{1\over 2}\vec\Delta, -\vec\kappa + {1\over
2}\vec\Delta\right)= {\cal F}\left(0.41{Q^2+m_V^2 \over W^2},\vec \kappa\right)
\exp\left(-{b_{3\Pom}\vec\Delta^2 \over 2}\right)\,,
\ee
where $b_{3\Pom}$ includes contributions from two-gluon formfactor 
of the proton and from the effective Pomeron trajectory,
as described in detail in \cite{review}. 
Although fixing the exact numerical value of the shift coefficient (0.41)
is beyond the log${1\over x}$ accuracy, its introduction is phenomenologically motivated
and, after all, it can be viewed as yet another parameter in our parametrization.
We did not try varying this parameter to obtain a better fit.  
The parametrizations for the forward unintegrated gluon density
were borrowed from \cite{in2000}.
Note that the $k_t$-factorization approach
itself does not require $Q^2$ to be large, and in the soft region, 
$Q^2 \lsim 1$ GeV$^2$, the words ``unintegrated gluon density'' should be understood
simply as an appropriately normalized Fourier transform of the dipole cross section.  

As for the radial wave function, we choose the simple Gaussian Ansatz WF normalized according
to (\ref{norm3}), with the only free parameter, the ``size'' $a$ of the wave function. 
In contrast to the vector meson case, dileptonic decay $\rho_3 \to e^+e^-$  
cannot proceed via one-photon annihilation and does not help us fix $a$.
However, appealing to the argument that the radial wave functions
of spin-orbital partners should be similar, we can take $a_{3} = a_{D}$,
the latter being extracted from $\Gamma(\rho(1700) \to e^+e^-)$.

\subsection{Level of accuracy anticipated}

Our experience with diffractive production of ground state vector mesons
within the same approach tells us that variation of the input parameters 
changes the absolute values of the cross sections by a factor of $\lsim 1.5$, 
while the accuracy for the observables that depend on the {\em ratios} 
of the amplitudes is even better \cite{review,phd}.
The principal source of uncertainty was found to be the final 
meson wave function, especially its density near the origin. 
The sensitivity of the results to particular parametrizations 
of the unintegrated gluon density presented in \cite{phd} 
(from both DGD2000 and DGD2002 sets of parametrizations) 
was found to be weak.

In the present case, the results are expected to be less stable with variation of input
due to the presence of various cancellations.
The main source of instability is the poorly known value of the $\rho_D$
dileptonic decay width.
The data available give $\Gamma(\rho_D \to e^+e^-) \sim 0.1$--0.6 keV
(assuming that $\rho(1700)$ is indeed $D$-wave $q\bar q$ state).
The possibility that $\rho(1700)$ has significant contributions from radially excited $q\bar q$
and from possible hydrid state, as well as taking into account extremely large
NLO corrections \cite{correction} for this decay, make the situation even less definite.

All the curves to be presented below were calculated for $\Gamma(\rho_D \to e^+e^-) = 0.14$ keV.
This value corresponds to the value of $\Gamma(e^+e^-)\cdot Br(\pi^+\pi^-) 
= 29 {}^{+16}_{-12}$ eV obtained in \cite{Kurdadze83}.
In order to see the effect of this input parameter, we calculated cross sections
both for $\rho_3$ and $\rho_D$ for the dileptonic decay width in the interval $0.14-0.7$ keV.
Increasing $\Gamma(\rho_D \to e^+e^-)$, we observed some suppression of the cross sections
at small $Q^2$ and their significant growth at $Q^2 \gsim 1$ GeV$^2$, 
especially in the case of $\sigma_T$. 
The effect is strong, and we conclude that 
the numerical results for the {\em absolute values} of the cross sections 
are trustable only within factors of $\sim 2$--3.

We stress, however, that variation of the input parameters produced 
absolutely the same shifts in the $\rho_D$ production cross sections. 
This should be expected, because the relation between the two mesons is dictated
primarily by the similarity of their radial distributions and by spin-angular relations
of type of (\ref{reason}). These relations are essentially insensitive to details
of the model, as long as we treat $\rho(1700)$ as a predominantly $q\bar q$ pair in
the $D$-wave state. We conclude therefore that the numerical values of our predictions for 
ratios between $\rho_3$ and $\rho_D$ are more stable, 
approximately within a factor of 1.5--2.

\subsection{$Q^2$ and $t$-dependence}

We calculated all the helicity amplitudes (\ref{i3})-(\ref{i0}) for spin-3 meson and compared 
its production rate with that of $\rho_D$ and $\rho_{1S}$. 
All the cross sections are calculated at $W=75$ GeV and are obtained
from numerical integration of the differential cross sections
within the region $0 < |t| < 1.05$ GeV$^2$. 

\begin{figure}[!htb]
   \centering
\includegraphics[width=10cm]{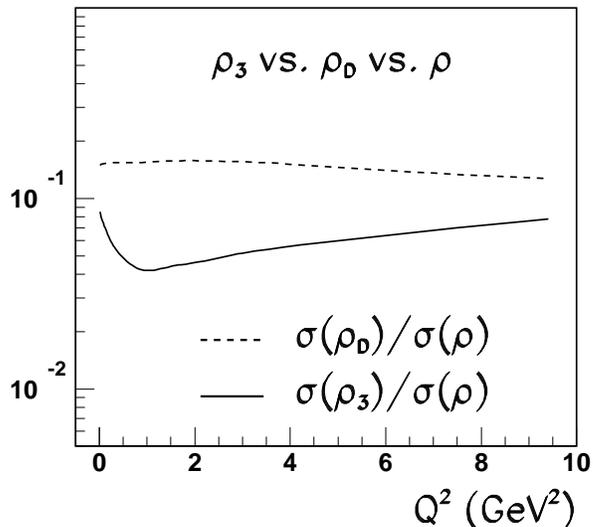}
\vspace{-1cm}
\caption{Predictions for the ratios of $\rho_3$ (solid line) and $\rho_D$ (dashed line) 
to $\rho$ production cross sections as a function of $Q^2$.}
   \label{fig-primes}
\end{figure}

Fig.~\ref{fig-primes} shows
the ratios of the excited to ground state meson cross sections 
$\sigma(\rho_3)/\sigma(\rho_{1S})$ and $\sigma(\rho_D)/\sigma(\rho_{1S})$.
Both ratios are an order of magnitude smaller than unity, and the $\rho_D$
cross section is noticeably larger than that of $\rho_3$, especially at $Q^2 \sim 1$ GeV$^2$.
At larger $Q^2$, the ratio $\sigma(\rho_D)/\sigma(\rho_3) \sim $ 2.
Thus, if one intends to extract $\rho(1700)$ properties 
from diffractively produced multipion states around 
invariant mass $\sim 1700$ MeV, one cannot neglect
 contamination by the $\rho_3$ state.

\begin{figure}[!htb]
   \centering
\includegraphics[width=14cm]{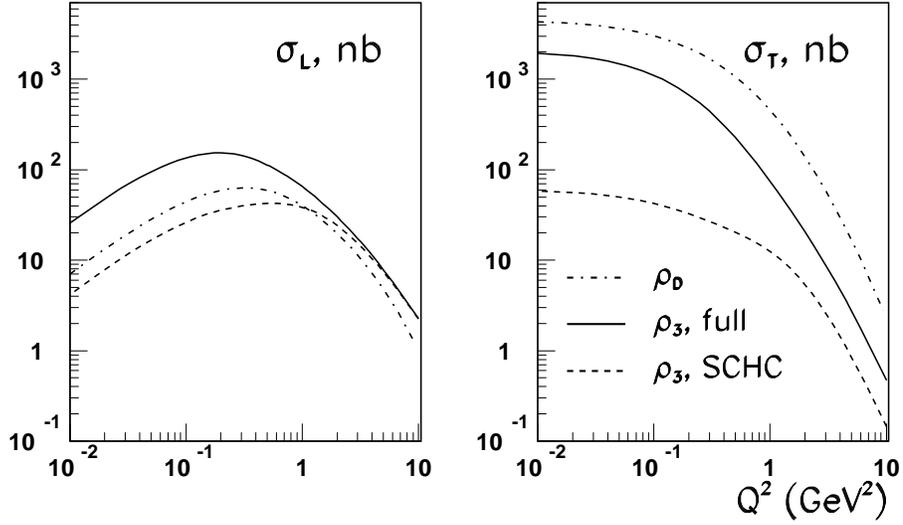}
\caption{Predictions for the longitudinal ({\em left plot}) and transverse ({\em right plot}) 
cross sections of $\rho_3$ (solid lines) and $\rho_D$ (dash-dotted lines) production.
The contribution to $\rho_3$ from SCHC amplitudes only is shown with
dashed lines.}
   \label{fig-sigmal-sigmat}
\end{figure}

\begin{figure}[!htb]
   \centering
\includegraphics[width=14cm]{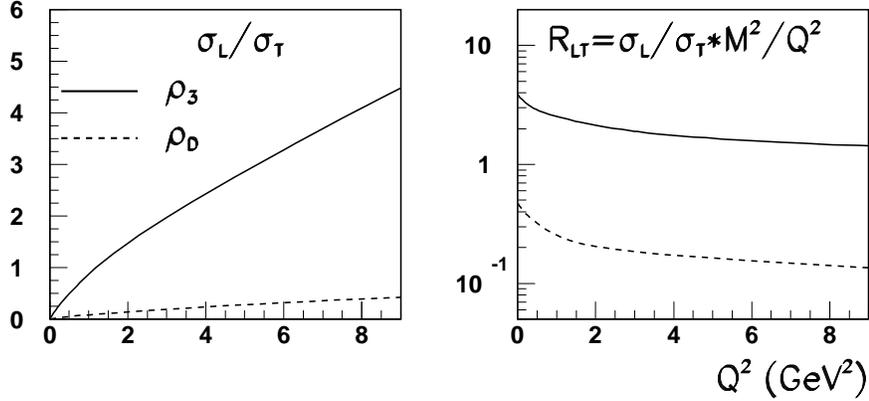}
\vspace{-1cm}
\caption{Predictions for the ratios of the longitudinal to transverse 
cross sections of $\rho_3$ (solid line) and $\rho_D$ (dashed line) production.
The left plot shows $R = \sigma_L/\sigma_T$, while the right plot shows reduced 
ratio $R_{LT} = \sigma_L/\sigma_T\cdot m_V^2/Q^2$.}
   \label{fig-ratiolt}
\end{figure}

Difference between $\rho_D$ and $\rho_3$ is better seen if one studies separately longitudinal 
and transverse cross sections, shown in Fig.~\ref{fig-sigmal-sigmat}.
Here, the solid and dash-dotted lines represent the $\rho_3$ and $\rho_D$ 
cross sections, respectively. 
In the case of $\rho_3$ we showed also with the dashed lines
the contributions of the SCHC amplitudes only. One clearly sees the domination of 
helicity violating amplitudes at small $Q^2$ in $\rho_3$ production. 
One can even state, on the basis of our calculations,
that $\rho_3$ production at small $Q^2$ {\em probes diffraction
in the regime of strong $s$-channel helicity violation}.
In the case of longitudinal cross section,
the contribution of SCHNC transitions becomes small 
at $Q^2 > 1$ GeV$^2$, since all such amplitudes are of higher twist. 
Helicity violation remains strong for transverse photons even
at large $Q^2$.

As mentioned above, in the case of
$D$-wave vector mesons one expects suppression of $\sigma_L$ but not in $\sigma_T$.
Indeed, our calculations show the domination of the $\rho_3$ over $\rho_D$
in $\sigma_L$ at small $Q^2 \lsim 1$ GeV$^2$, while in $\sigma_T$ the $\rho_D$ 
cross section is noticeably larger than $\rho_3$ everywhere.  
This is in a qualitative agreement with twist analysis result (\ref{s31twist}).

Such a different behavior of $\rho_D$ and $\rho_3$ can be seen also in
plots of ratio $\sigma_L/\sigma_T$ as a function of $Q^2$, shown in 
Fig.~\ref{fig-ratiolt}. Here, on the left plot, we showed this ratio for $\rho_3$ 
and $\rho_D$, while on the right plot,
we showed reduced ratios $R_{LT} = {\sigma_L \over \sigma_T}\cdot {m_i^2 \over Q^2}$,
where $m_i$ is the mass of the corresponding meson. $R_{LT}$ is small for $\rho_D$ 
and relatively large for $\rho_3$, as was expected from the twist analysis (\ref{rlt}).
We note that the region $Q^2 \sim 1$ GeV$^2$ is particularly suitable for distinguishing
among various $\rho$ states.

\begin{figure}[!htb]
   \centering
\includegraphics[width=14cm]{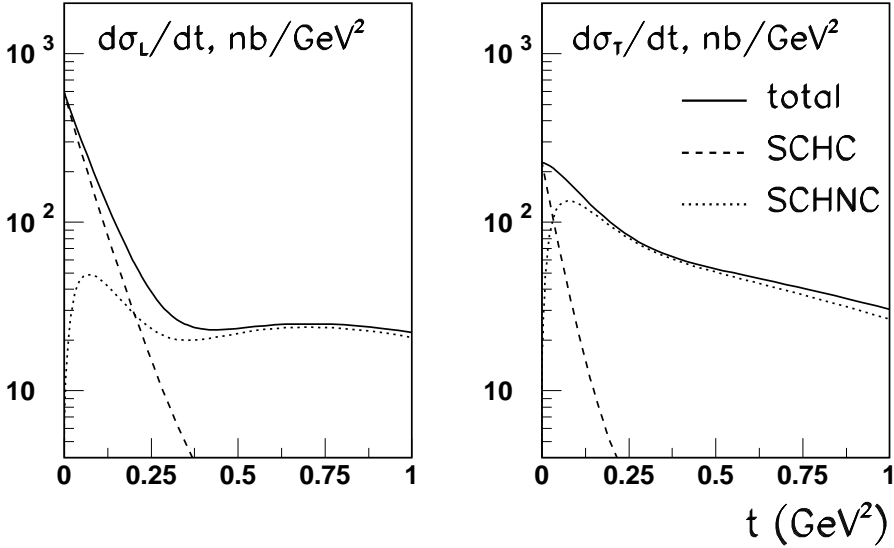}
\vspace{-1cm}
\caption{Differential cross sections $d\sigma_L/d|t|$ (left plot) 
and $d\sigma_T/d|t|$ (right plot) for $\rho_3$ at $Q^2 = 1$ GeV$^2$ 
as functions of $|t|$.
The dashed and dotted lines show contributions from helicity conserving 
and helicity violating amplitudes, respectively; the solid lines
show their sum. }
   \label{fig-t}
\end{figure}

The role of helicity violating amplitudes can be seen also 
in the $|t|$-distributions shown in Fig.~\ref{fig-t} for $Q^2=1$ GeV$^2$ 
separately for longitudinal and transverse cross sections.
The dashed and dotted lines show the SCHC contributions only,
while the solid lines show their sums.
As can be expected, the large contributions of SCHNC amplitudes come from
the entire $t$-interval shown, while the SCHC amplitudes are strong only 
within a narrow forward cone.

\subsection{Energy dependence}

We checked also the energy dependence of the $\rho_3$ production cross section,
which we parametrized with a simple power law $\sigma(\rho_3) \propto W^\delta$.
General expectations, driven by the ground state production experience \cite{review}, 
 are that at small $Q^2$ this exponent should be a small number,
and one speaks usually of ``soft Pomeron'', while in the presence of hard scale
it should grow up to $\delta \sim 1$, corresponding to the ``hard Pomeron''.

For a particular example, we used HERA kinematics and studied
the energy behavior of the cross section within the range $W = $ 50--200 GeV.
We found that at small-to-moderate $Q^2$, the $\rho_3$ production cross section
(both longitudinal and transverse) 
slightly {\em decreases} with energy rise, with typical $\delta \sim -0.1$ to $-0.2$.
It is only at $Q^2 \gsim 5$ GeV$^2$ that $\delta$ becomes positive,
and it is always smaller than the value of $\delta$ for the corresponding ground state
vector meson.

This decrease is naturally understood in the Regge picture of the Pomeron
exchange. The differential cross section at non-zero $t$ behaves roughly as
$$
{d\sigma \over dt} \propto W^{\delta(t)}\,;\quad
\delta(t) = 4[\alpha_\Pom(t)-1] 
\approx 4[\alpha_\Pom(0)-1 - \alpha^\prime_{eff}\cdot |t|)]\,.
$$
The value of the effective Pomeron intercept $\alpha_\Pom(0)$ 
depends on $Q^2$ and comes directly from the parametrizations
of the forward unintegrated gluon density, see \cite{in2000}.
At small $Q^2$ it is about $\alpha_\Pom(0)-1 \sim 0.08$, 
and starts noticeably growing only at $Q^2 \gsim$ 2--3 GeV$^2$.  
The effective slope of the Pomeron trajectory is 
$\alpha^\prime_{eff}\approx 0.12$ GeV$^{-2}$ with very 
marginal $Q^2$ dependence. (Note that it differs
from the fixed input parameter $\alpha^\prime_{eff}\approx 0.25$ GeV$^{-2}$
used in our calculations due to anti-shrinkage effects discussed in 
detail in \cite{antishrinkage}.) Thus, the effective exponent
of the energy dependence of the integrated cross section 
$$
\delta \approx 4[\alpha_\Pom(0)-1 - \alpha^\prime_{eff}\cdot \langle |t| \rangle]
$$
is governed not only by $\alpha_\Pom(0)$, but also by the typical momentum transfers
$ \langle |t| \rangle$ involved.

In the production of ground state vector mesons dominant contribution
comes from SCHC helicity amplitudes, which are
concentrated within the forward cone $|t|\lsim 0.1$ GeV$^2$.
In the present case, as Fig.~\ref{fig-t} shows vividly,
the range of important values of $|t|$ spans up to $0.5$--1 GeV$^2$.
One sees that due to such high values of $|t|$ involved 
the energy rise exponent $\delta$
at small $Q^2$ can easily become negative.   

\section{Discussion}

\subsection{What dipole sizes are probed in $\rho_3$ photoproduction?}

The $S$-wave $q\bar q$ state is naturally orthogonal to the $D$-wave
$q\bar q$ state. For example, if one attempts to calculate the inelastic 
$\rho_{S} \to \rho_{D}$ formfactor at zero momentum transfer, 
one finds
$$
{\cal M} \propto \int {dz \over z(1-z)} d^2\vec k\, \psi^*_D(\bp^2) \psi_S(\bp^2)
(2k_z^2 - \vec k^2) = \int {4 \over M} d^3\bp\, \psi^*_D(\bp^2) \psi_S(\bp^2)
(2k_z^2 - \vec k^2) = 0\,.
$$
The presence of the quadrupole combination $2k_z^2 - \vec k^2$
makes the amplitude zero, as long as all other factors
under the integral are spherically symmetric. 

In the photoproduction, the wave function of the initial photon 
is not spherically symmetric, but one can still appeal 
to the vector dominance model arguments and rewrite
the $\rho_3$ photoproduction amplitude as
\be
\langle \gamma |\sigma(r)|\rho_3\rangle \propto 
g_{\gamma \rho}\cdot \langle \rho |\sigma(r)|\rho_3\rangle\,.
\ee 
One can suspect that a similar orthogonality should be at work here,
when one considers a forward SCHC amplitude 
at large $q\bar q$ dipole sizes, where the dipole cross section 
$\sigma(r) \to$ const so that all other factors seemingly become
spherically symmetric.
If this were the case, it would mean that the $\rho_3$
photoproduction receives little contribution from large dipoles
and is a ``harder'' process than the $\rho$ photoproduction.
However, this is not the case. 
The most essential difference between the $\rho p \to \rho_3 p$ amplitude and 
(\ref{ampl}) is the replacement of the photon wave function:
\be
\Phi_2 \to \Psi_2 = {1\over z(1-z)}\left[2\psi_S(k_z^2 + \vec k^2)
- \psi_S(k_z^2 + (\vec k+\vec\kappa)^2) - \psi_S(k_z^2 + (\vec k-\vec\kappa)^2)\right]\,,
\ee
together with a similar replacement of $\vec\Phi_1$.
For clarity, we explicitly presented the spherically symmetric quantity 
$\bp^2$ as $k_z^2 + \vec k^2$.
At small values of $\vec\kappa^2$, which correspond to large dipole sizes,
one obtains:
\be
 \Psi_2 \approx - 2\vec\kappa^2 \left[\psi_S^\prime(\bp^2) + \vec k^2 
\psi_S^{\prime\prime}(\bp^2)\right]\,,\label{psismall}
\ee
where derivatives of the radial wave function are taken in respect to 
$\bp^2$. 

The result (\ref{psismall}) explicitly lacks spherical symmetry.
Thus, even if one uses the leading twist contributions, pretending that
the $\rho \to \rho_3$ transition 
is well approximated by non-relativistic expressions,
one still gets
\be
\int d^3\bp  \left[\psi_S^\prime(\bp^2) + \vec k^2 
\psi_S^{\prime\prime}(\bp^2)\right]\,\psi_3 \cdot (2k_z^2 - \vec k^2) \not = 0\,.
\ee 
This result is rather natural. The characteristic
feature of the high-energy collision is presence of a preferred
direction: that of the proton's momentum in the vector meson rest frame.
The transverse and longitudinal dynamics of the quark loop now differ,
and this leads, in particular, to circulation of purely transverse
momentum $\vec\kappa$ in the loop, see (\ref{psismall}),
which breaks the spherical symmetry.  Taking into account the Fermi motion
makes the expression to be integrated even less symmetric.

\begin{figure}[!htb]
   \centering
\includegraphics[width=12cm]{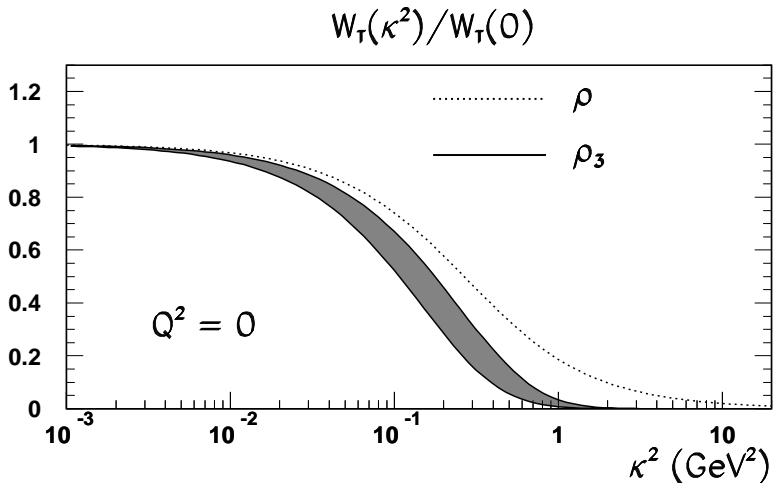}
\vspace{-1cm}
\caption{The mapping functions $W_T(\vec\kappa^2)$ normalized 
to the $\vec \kappa^2 = 0$ point for the $\rho_3$ (solid lines) 
and $\rho_{1S}$ (dotted line) photoproduction.
The shaded area shows the distribution of the results due to 
the decay width variation $\Gamma(\rho_D \to e^+e^-) = 0.14$--0.7 keV.}
   \label{fig-w}
\end{figure}

It is useful to check explicitly that large dipoles indeed contribute substantially
to $\rho_3$ production amplitude. In order to test this numerically,
we change the order of integration in (\ref{ampl}) 
and represent the forward $T \to T$ amplitude as
\be
{1\over W^2} Im\, A_{+1;+1} =  \int {d\vec \kappa^2 \over \vec \kappa^2}
{\cal F}(x_1,x_2,\vec \kappa,0) \cdot W_T(\vec \kappa^2)\,.  \label{amplkap}
\ee
One expects $W_T(\vec \kappa^2)$ to have the ``smoothed step function'' shape:
it should be approximately constant at small $\vec \kappa^2$ up to some
value $\overline{Q_T}^2$, and should decrease quickly as $\vec\kappa^2$ 
passes this value, see \cite{hardscale} for details of this analysis
for the ground state $\rho$ production.
This function ``cuts out'' the important range of gluon momenta,
and determines thus the important range of the color dipole sizes.
The effect of orthogonality --- if it were present --- would 
appear in $W_T(\vec\kappa^2)$ as a small-$\vec\kappa^2$ suppression.

In Fig.~\ref{fig-w} we show $W_T(\vec\kappa^2)$ 
normalized to the $\vec\kappa^2=0$ point for the $\rho_3$ and $\rho_{1S}$
photoproduction with solid and dotted lines, respectively.
The shaded area corresponds to scattering of the results
due to variation of the decay width 
$\Gamma(\rho_D \to e^+e^-) = 0.14$--0.7 keV.
We observe no small-$\vec\kappa^2$ suppression for $\rho_3$. 
The $\vec\kappa^2$ value where $W(\vec\kappa^2)/W(0)$ hits $1/2$ 
is $\vec \kappa^2 \approx 0.1$--0.2 GeV$^2$ for the $\rho_3$,
and is noticeably smaller than the corresponding value of 0.27 GeV$^2$ 
in the $\rho_{1S}$ production.
One sees that the $\rho_3$ production is indeed a softer process,
and the typical dipole sizes probed are $\sim 1.2\div 1.5$ time larger 
than in the ground state VM production.

\subsection{Comparison with experimental data available}

The OMEGA Collaboration at CERN measured the cross section of 
diffractive photoproduction of $\rho_3(1690)$ (known then as $g(1690)$ meson) 
via the $a_2(1320)\pi$ subsample of the $\eta\pi^+\pi^-$ diffractive 
final state events \cite{omega1986}.
The cross section of $\gamma p \to \rho_3(1690)p \to a_2(1230)\pi p$
was found to be $97 \pm 28 \pm 21$ nb, which allows one to roughly estimate the $\rho_3$
production cross section as $\sigma(\gamma p \to \rho_3(1690)p) \sim$ 200--300 nb.
This result is about 5--10 times below our photoproduction predictions,
which we think is not very bad discrepancy, taking into account 
expected level of accuracy in the soft region. 
Indeed, most of this cross section
we predicte to be due to SCHNC, especially the double-flip, transitions.
Its magnitude in the soft region was predicted by our calculation
to be rather large even for ground state vector mesons, 
but so far has been poorly known from experiment.
We think that upon understanding better the role of SCHNC at small $Q^2$
with the aid of modern experiments, we can improved the accuracy of our predictions.
We expect, however, that our conclusion of the strong violation of $s$-channel helicity
violation probed in $\rho_3$ will survive such an upgrade.
In addition, our cross sections correspond to integration within $0 < |t| < 1.05$ GeV$^2$;
the results will change noticeably if one selects another $t$-interval.

The OMEGA Collaboration also measured the photoproduction cross section of $\rho'$.
The original data were reanalysed in terms of $\rho(1450)$ and $\rho(1700)$ separately
in \cite{tworhoprimes} yielding $\sigma(\rho(1700)) \sim 500$ nb. Thus, our result
$\sigma(\rho(1700))/\sigma(\rho_3) \sim 3$ at the photoproduction limit is roughly 
consistent with experiment. Finally, comparing the $\rho_3$ and the ground state $\rho$ 
photoproduction cross sections, we note that our result $\sigma_3/\sigma_{1S} \sim 0.1$
is again not very far from experimental value of 0.02--0.03.

\subsection{Comments on experimental possibilities}

The experimental analysis of diffractive production of spin-3 resonances,
and in particular, the strategy of $\rho_3(1690)/\rho(1700)$ separation, 
will depend on the statistics available and the final state chosen. 

Should one have the luxury of high statistics, one can do the partial wave analysis
or select some particular final states, in which one of the two states would dominate.
An example of this approach is just the OMEGA Collaboration observation of the $\rho_3$ in the
$a_2(1320)\pi\to \eta\pi^+\pi^-$ final state.
If the statistics does not allow for such angular dependence or final state analysis, 
one then should look for distinctions in the production of these mesons.
In the view of our results, it is tempting to make use of ratios $\sigma_L/\sigma_T$,
which are dramatically different for $\rho_3$ and $\rho_D$, especially 
in the small-to-moderate $Q^2$ region. 

One possibility to separate $\sigma_L$ and $\sigma_T$
is given by the Rosenbluth method. 
It will require several runs at different lepton beam energies
and might seem impractical at high energies.
The second possibility could be to do a baby-version of PWA and to 
study angular correlations in final state hadrons. 
For example, if both mesons discussed are observed in $\pi^+\pi^-$ 
(the corresponding branching ratios are not dominant, 
but still sizable), one could study the single-differential 
angular distribution $W(\cos\theta)$. 
This measurement will give spin density matrix element $r^{04}_{00}$, from which
one recovers $\sigma_L/\sigma_T$. Alternatively, one can search for a similarly revealing
angular dependence in $4\pi$ final states, the dominant decay channel of both mesons.

Another issue, which requires taking into account the $\rho_3$ meson,
is the recent observation of a narrow dip structure in diffractively photoproduced 
$3\pi^+3\pi^-$ states at $M_{6\pi}\approx 1.9$ GeV in Fermilab E687 experiment \cite{E687-6pi}. 
Although the detailed mechanism of its appearance remains unsettled,
the very recent analysis \cite{E687-6pianalysis} sees it as a result of interplay of several
resonances with $J^{PC} = 1^{--}$ (including $\rho(1700)$) and a background. 
This analysis was explicitly based on the vector dominance idea 
and explicitly uses the assumption that the $6\pi$ spectra
in $e^+e^-$ annihilation and in diffractive photoproduction are essentially the same 
(apart from kinematical factors). The results presented here clearly show
that this is a risky assumption. The $\rho_3$ meson does not couple to the single virtual photon,
yet it should be produced diffractively at a rate comparable to that of $\rho(1700)$.   
Although it cannot produce any interference pattern with $J=1$ states, 
its own contribution can affect the results of a very delicate analysis of \cite{E687-6pianalysis}. 

\section{Conclusions}

We calculated the cross section of the exclusive production of $J^{PC}=3^{--}$ mesons 
in diffractive DIS within the $k_t$-factorization approach. The results were compared
with the cross section of the $D$-wave state vector meson 
of the same quarkonium. We exemplified the general expressions with a detailed
numerical study of the $\rho$ system, where the $\rho_3(1690)$ state is almost
degenerate with the $\rho(1700)$ meson, whose structure is arguably dominated by 
the $q\bar q$ pair in the $D$ wave. 

The absolute values of the cross sections suffer from uncertainties
of the input parameters, in particular, of the $\rho_D \to e^+e^-$ decay width,
and we can be sure only in the order of magnitude of these results.
However, in what concerns the relative production rates of $\rho_3$ and $\rho_D$,
our conclusions are much more certain. Our results allow us to formulate
the following predictions, which are stable against variations of the model parameters:
\begin{itemize}
\item In typical HERA kinematics, the ratio of production cross sections 
taken at equal $Q^2$ is $\sigma(\rho_D)/\sigma(\rho_3) \approx$ 3--5 at small $Q^2$,
decreasing to $\approx 2$ at larger $Q^2$. 
Thus, when extracting the properties of $\rho(1700)$ from multipion final states,
one cannot simply neglect the $\rho_3$ contribution.
\item $\rho_3$ and $\rho_D$ show completely different patterns 
in $\sigma_L$-$\sigma_T$ decomposition: $\rho_3$ dominates in the longitudinal cross section,  
while $\rho_D$ dominates in the transverse cross section. 
The ratios $R = \sigma_L/\sigma_T$ for $\rho_3$ and $\rho_D$ differ 
by more than one order of magnitude. This dramatic difference
can be traced back to spin-angular properties of these two mesons, see (\ref{reason}).
\item The role of $s$-channel helicity violating amplitudes is extremely important,
especially in the transverse cross section. At small-to-moderate $Q^2$
the helicity violating amplitudes even dominate over the SCHC ones.
Thus, production of $\rho_3$ offers an interesting possibility to study diffraction in
the regime of strong $s$-channel helicity violation.
\item Due to large color dipole sizes probed in the $\rho_3$ production
and large region of relevant momentum transfers, $|t|\lsim 1$ GeV$^2$,
the energy dependence of $\rho_3$ production cross section
is less steep than in the case of $\rho$. At small $Q^2$, one might even observe
decrease of $\rho_3$ cross section with energy rise.  
\end{itemize}
We find no surprise in numerical stability of the first two conclusions,
since they are essentially driven by very basic relations: similarity of radial wave functions
for $\rho_3$ and $\rho_D$, the spin-angular composition of these mesons, see (\ref{reason}),
and quadrupole suppression of the leading contributions 
in SCHC amplitudes (\ref{i00}) and (\ref{ipp}).  

In addition, confronting our predictions for photoproduction with the fixed target data available
and observing them to agree within the anticipated accuracy inspires hope 
that we grasp the essential physics of this reaction in our approach.
We are looking forward to seeing experimental checks of our predictions.\\

It is our pleasure to acknowledge fruitful discussions with Etienne Burtin, Roberto Fiore, 
Nicole d'Hose, Dima Ivanov, Kolya Nikolaev, Alessandro Papa, Andrzej Sandacz.
The work of I.P.I. is supported by the INFN Fellowship, and partly by INTAS and grants 
RFBR 05-02-16211 and NSh-2339.2003.2.

\end{document}